\documentclass[aps,prb,showpacs,amsmath,amssymb,twocolumn,superscriptaddress]{revtex4-1}
\usepackage{color}
\usepackage{amsmath}
\usepackage{graphicx}

\begin{document}

%\pdfoutput=1
\newcommand{\bn}{{\bf n}}
\newcommand{\bp}{{\bf p}}   
\newcommand{\br}{{\bf r}}
\newcommand{\bk}{{\bf k}}
\newcommand{\bv}{{\bf v}}
\newcommand{\brho}{{\bm{\rho}}}
\newcommand{\bj}{{\bf j}}
\newcommand{\wk}{\omega_{\bf k}}
\newcommand{\nk}{n_{\bf k}}
\newcommand{\eps}{\varepsilon}
\newcommand{\la}{\langle}
\newcommand{\ra}{\rangle}
\newcommand{\be}{\begin{eqnarray}}
\newcommand{\ee}{\end{eqnarray}}
\newcommand{\intl}{\int\limits_{-\infty}^{\infty}}
\newcommand{\dE}{\delta{\cal E}^{ext}}
\newcommand{\SE}{S_{\cal E}^{ext}}
\newcommand{\dsp}{\displaystyle}
\newcommand{\phit}{\varphi_{\tau}}
\newcommand{\p}{\varphi}
\newcommand{\cL}{{\cal L}}
\newcommand{\dphi}{\delta\varphi}
\newcommand{\dbj}{\delta{\bf j}}

\newcommand{\rred}[1]{{#1}}
\newcommand{\skp}[1]{{#1}}
\newcommand{\bblue}{}

\title{Charge and spin current in a quasi-one-dimensional quantum wire with spin-orbit coupling}

\author{K.~E.~Nagaev}
\affiliation{Kotelnikov Institute of Radioengineering and Electronics, Mokhovaya 11-7, Moscow, 125009 Russia}
\affiliation{Moscow Institute of Physics and Technology, Institutsky per. 9, Dolgoprudny, 141700 Russia}

\author{A.~S.~Goremykina}
\affiliation{Kotelnikov Institute of Radioengineering and Electronics, Mokhovaya 11-7, Moscow, 125009 Russia}
\affiliation{Moscow Institute of Physics and Technology, Institutsky per. 9, Dolgoprudny, 141700 Russia}

\date{\today}

\begin{abstract}
We show that Rashba spin-orbit coupling may result in an energy gap in the spectrum of electrons in a two-mode quantum wire
if a suitable confining potential is chosen. This leads to a dip in the conductance and a spike in the spin current at 
the corresponding position of the Fermi level. Therefore one may control the charge and spin currents by means of 
electrostatic gates without using magnetic field or magnetic materials.
\end{abstract}

\maketitle

\section{Introduction}

Spintronics, or spin electronics, involves the study of active control and manipulation of spin degrees
of freedom in solid-state systems and is a rapidly growing field of science.\cite{Zutic04} The key purpose of these studies is 
the generation, control, and manipulation of spin-polarized currents. A useful tool for achieving this goal is the spin-orbit 
interaction, which couples the spin of an electron with its spatial motion in a presence of a certain asymmetry of the conductor.
For example, Rashba spin-orbit interaction is due to a lack of inversion symmetry in semiconductor heterostructures such as InAs or 
GaAs.\cite{Rashba60} The advantage of this type of interaction is that it can be tuned by means of electrostatic 
gates.\cite{Nitta97,Engels97} 

In truly single-mode quantum channels, spin-orbit interaction alone neither changes the electric current nor results in a spin current if no magnetic field or magnetic materials are involved. In this case, spin-orbit interaction does not change the energy-band topology and can be simply eliminated 
from the Hamiltonian by means of a unitary transform.\cite{Levitov03} 

A prototypical scheme of a spin field-effect transistor based on Rashba interaction and single-mode ballistic channel with ferromagnetic electrodes
was proposed by Datta and Das\cite{Datta90} more than two decades ago. Recently, such a device was experimentally realized.\cite{Koo09}
 
The current in a single-mode quantum channel also depends on the spin-orbit interaction if a magnetic field is applied parallel to the channel or
normally to the plane of the heterostructure (i.e. in the direction of Rashba field).\cite{Pershin04} The interplay of the spin-orbit interaction with 
magnetic field significantly modifies the band structure and produces an energy gap in the spectrum together with additional subband extrema. This results in a decrease in the charge current and a net spin current as the Fermi level passes through the gap. These effects were recently experimentally observed by Quay et al.\cite{Quay10}

Many authors studied spin and charge transport in multimode quantum channels in the absence of magnetic field or magnetic ordering. Governale and Z\"ulicke\cite{Governale02} considered a long channel with parabolic confinement potential and took into account the mixing of different 
transverse-quantization subbands by the spin-orbit interaction. This mixing results in an asymmetric distortion of the dispersion curves but does not open any gaps in the spectrum. As a consequence, the spin-orbit interaction in a presence of voltage drop across the channel results in a spin accumulation inside the channel but does not lead to a spin current or deviations from the standard conductance quantization. There is also a number of numerical calculations of the spin current,\cite{Eto05,Zhai07,Liu07,Zhai08} but these papers deal with stepwise constrictions and the results are obscured by the interference effects. A more realistic geometry of a saddle-point contact in two-dimensional potential landscape was considered in Ref.~\onlinecite{Sablikov10}, but the Rashba interaction was taken  into account there as a perturbation. In Refs. \onlinecite{Sanchez06} and
\onlinecite{Gelabert10}, a quasi-one-dimensional wire with localized region of Rashba interaction was considered and nonzero spin current was predicted
for sufficiently sharp boundaries of the region. Unusual trajectories were revealed
by Silvestrov and Mishchenko\cite{Silvestrov06} within the quasiclassical
approach to exist near these regions. However in all the above papers, the spin current and the deviations from perfect conductance quantization are related with the mixing of subbands in the transition areas between the quantum contact and reservoirs by spin-orbit interaction and crucially depend on the geometry and properties of these regions. It is hard to see any general regularities concerning the magnitude of the effect.

In this paper, we propose a mechanism of spin current generation that relies on the energy band structure deep in the wire rather than on the reflection effects in the transition areas and leads to  100\% spin-polarized current at definite positions of the Fermi level. This mechanism is reminiscent of the one in Ref. \onlinecite{Pershin04} but requires no magnetic field.

\section{The model}

Consider a quasi-one-dimensional conducting channel formed in two-dimensional electron gas by means of electrostatic gates. The transition between the reservoirs and the channel is assumed to be adiabatic, and the length of the channel is much larger than that of the transition regions. We assume that the Rashba  spin-orbit interaction is present in the channel but absent in the reservoirs, so the spin current through the system is well-defined.  The Hamiltonian of the system is of the form 
\begin{multline}\label{H}
\hat{H}=\frac{\hat{p}^2_x}{2m}
+\frac{\hat{p}^2_z}{2m}+U(x,z)
\\
+\frac{\alpha(x)}{\hbar}
 \left(\hat{p}_x\hat{\sigma}_z-\hat{p}_z\hat{\sigma}_x\right)
-\frac{i}{2}\,\frac{\partial\alpha}{\partial x}\,\hat{\sigma}_z,
\end{multline}
where $U(x,z)$ is the confining potential and  $\alpha(x)$ is the parameter of spin-orbit coupling. Both quantities are smooth functions of the 
longitudinal coordinate $x$ that are constant almost throughout the whole length of channel and vanish in the reservoirs. It is now straightforward
to make use of the adiabatic approximation and introduce a complete set of of eigenfunctions $\varphi_{n}(x,z)$ and eigenenergies $\eps_{n}$  
corresponding to the transverse motion of electrons in the $z$ direction. This leads to a set of coupled equations of the form
\begin{multline}
 \left[
  \frac{\hat{p}_x^2}{2m} + \frac{\alpha(x)}{\hbar}\,\hat{\sigma}_z\,\hat{p}_x - \frac{i}{2}\,\frac{d\alpha}{dx}\,\hat{\sigma}_z + \eps_m
 \right] \bar{\psi}_m(x) 
\\
 -\frac{\alpha(x)}{\hbar}\,\hat\sigma_x \sum_n \la m|p_z|n \ra\,\bar{\psi}_n(x) = \eps\,\bar\psi_m(x)
 \label{Schr-multi}
\end{multline}
for the spinors $\bar{\psi}_n = (u_n,\, v_n)^T$ that describe the longitudinal dependence of the spin-up and spin-down amplitudes of wave-function 
in the $n$-th transverse quantum mode. 

\begin{figure}[t]
 \includegraphics[width=8.5cm]{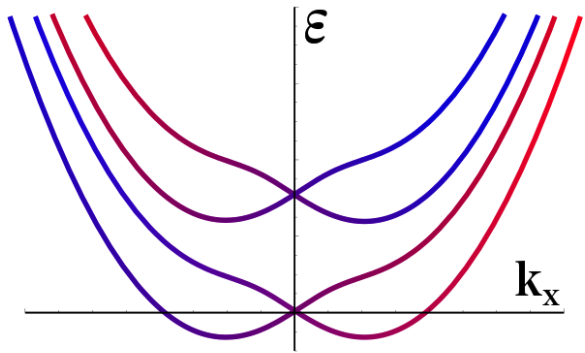}
 \caption{\label{fig1} Dispersion curves for the two lowest subbands of a quantum wire with spin-orbit interaction and parabolic
          transverse confinement. The dispersion curves are distorted by level crossing but exhibit no local maxima.The color of the curve
          designates the dominant spin projection.}
\end{figure}

If the matrix elements of transverse momentum %$\la m|p_z|n \ra$  
between different modes $\p_m$ and $\p_n$ were zero, the twofold spin degeneracy of these modes would be lifted by the spin-orbit interaction and one should see two sets of parabolic dispersion curves shifted along the $k_x$ axis
that would correspond to the two possible projections of spin on the $z$ axis. The curves of each set would have minima at $k_x=\pm 2m\alpha/\hbar^2$ and intersect without affecting each other.

Nonzero matrix elements $\la m|p_z|n \ra$ result in anticrossing of the dispersion curves with different $n$ and spin projection and lead to an asymmetric distortion of them (see Fig. \ref{fig1}).
However this does not give rise to new maxima and minima in these curves for the case of standard parabolic confining potential.\cite{Moroz99}
The reason is  that the levels of transverse quantization are evenly spaced and it is impossible to isolate a pair of them with a small separation. In other words, the vertical separation of anticrossing curves is too large as compared with their horizontal shifts.

The failure of the approximate two-band model that predicts a nonmonotonic behavior of the curves may be understood as follows.
In the absence of band mixing, the two curves corresponding to two subsequent transverse-quantization levels and different spin projection would cross at $k_x= \Delta\eps/2\alpha$,
where $\Delta\eps = \eps_{n+1} - \eps_n$. To form a maximum, the crossing branches of these curves should have different signs of slope $k_x+ 2m\alpha/\hbar^2>0$
and $k_x - 2m\alpha/\hbar^2 < 0$ at the intersection point, which results in a condition $\Delta\eps < 4m\alpha^2/\hbar^2$. On the other hand, the band mixing term $\alpha\la n+1|p_z|n\ra/\hbar$ would
lead to the splitting of the curves at the crossing point of the order of 
$\Omega \sim \sqrt{m\Delta\eps}\,\alpha/\hbar$. The two-band model is justified only if
$\Omega \ll \Delta\eps$, i.e. $\Delta\eps \gg m\alpha^2/\hbar^2$, which is incompatible with the previous condition. Exact calculations\cite{Governale02}
show that all the dispersion curves have only one minimum and hence the dependence of the conductance of the channel on the Fermi energy exhibits only the conventional $2e^2/h$ steps, while the spin current is absent.

\begin{figure}[t]
 \includegraphics[width=8.5cm]{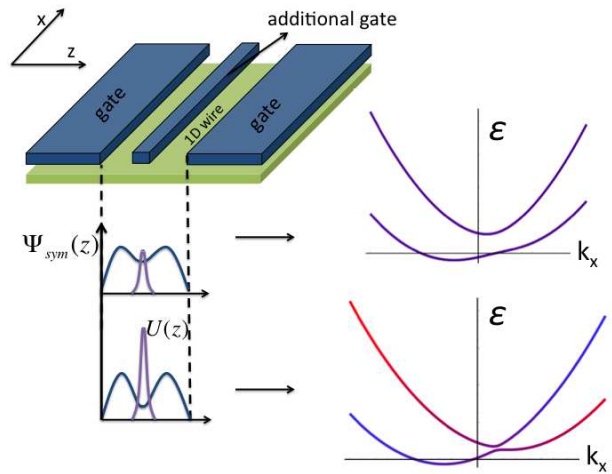}
 \caption{\label{fig2} The system under consideration. The current flows in the $x$ direction, and the negative voltage at the additional middle gate
 changes the confining potential from one-well to a double-well shape. As the negative voltage increases, a maximum appears in the lower
 dispersion curve.}
\end{figure}

\begin{figure}[t]
 \includegraphics[width=8.5cm]{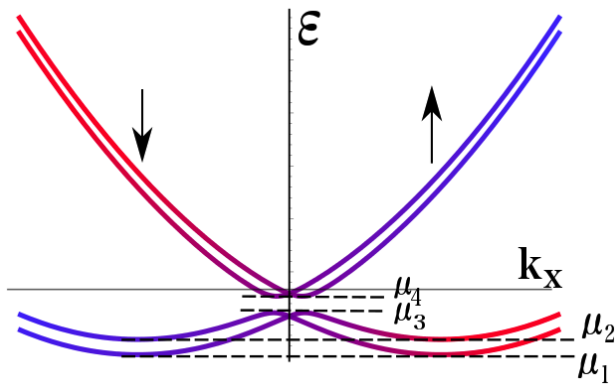}
 \caption{\label{fig3} Dispersion curves for a pair of closely
 spaced energy levels with small matrix element of transverse momentum in a quantum wire with spin-orbit interaction. The level crossing
 results in appearance of local maxima in the lower curves.}
\end{figure}

Things become different if the confinement is nonparabolic. Consider, e. g., the system in which  $U(z)$  has the shape of a double potential well. Such a potential may be formed by means of a negatively biased central gate on top of the quantum wire (see Fig. \ref{fig2}). In the case of a high impenetrable barrier between the wells, each of them would possess the same set of energy levels, so the levels of the whole system would be doubly degenerate. A finite tunneling through the barrier lifts the degeneracy, and therefore one gets a set of pairs of levels with very small spacings inside the pair.
In this case, the two-band model is justified and taking into account only the two lowest levels, one obtains the following expression for the  resulting dispersion curves
\begin{multline}
 \varepsilon = \frac{\hbar k_x^2}{2m} + \frac{\eps_1+\eps_2}{2} 
\\
 \pm \frac{1}{2}\sqrt{(\eps_2-\eps_1\pm 2 \alpha k_x)^2+4 |p_{12}|^2 \alpha^2/\hbar^2},
 \label{E(k)}
\end{multline}
where $p_{12}=\la 1|p_z|2\ra$. The upper sign at $2\alpha k_x$ under the square root corresponds to the mixture of 
$|1\uparrow\ra$ and $|2\downarrow\ra$ states, and the lower sign  corresponds to the mixture of $|1\downarrow\ra$ and $|2\uparrow\ra$ states.
The two pairs of the resulting curves are symmetric with respect to $k_x=0$. The lower dispersion curve may have one or two minima as a function 
of $k_x$ depending on the relations between $\Delta\eps = \eps_2 - \eps_1$, $p_{12}$, and $\alpha$ (see Fig. \ref{fig3}). The upper
minimum disappears by merging with the local maximum, i.~e. when the points where $d\eps/dk_x =0$ and  $d^2\eps/dk_x^2=0$ coincide. Therefore it  
follows from Eq.~(\ref{E(k)}) that the second minimum exists if
\begin{equation}
\left| \frac{\hbar p_{12}}{m\alpha}\right|^{2/3}+\left( \frac{\hbar^2\Delta \eps}{2m\alpha^2}\right)^{2/3}<1.
\label{inequality}
\end{equation}
Apparently one can meet this condition by making the overlap of the wave functions in the two wells sufficiently small. For example, 
if a square quantum well with infinitely high external walls is symmetrically cut by a $\delta$-like barrier in the middle, both $\Delta\eps$
and $|p_{12}|$ are inversely proportional to the effective strength of the barrier $k_0$.

\section{The conductance}

The existence of a local maximum in the lower pair of the dispersion curves leads to significant changes in the conductance of the wire. If the Fermi level lies between the lower and upper minima $\mu_1$ and $\mu_2$ in the dispersion curves (see Fig. \ref{fig3}), it intersects two branches with positive (negative) group velocity that correspond to the two different spin projections in the $z$ direction, and the conductance is $2e^2/h$, while the spin current is absent. If the Fermi level lies between the upper minimum $\mu_2$ and the local maximum $\mu_3$ in the lower curves or above the minimum in the two upper dispersion curves $\mu_4$, it intersect 
two branches with positive (negative) group velocity and one spin projection and two branches with positive (negative) group velocity with the other spin projection. This results in the $4e^2/h$ conductance and yields no spin current. However if the Fermi level falls within the gap between the local maximum 
$\mu_3$ in the lower curves and the minimum $\mu_4$ in the upper curves, it intersect two branches with positive group velocities and one spin projection and two branches
with negative velocities and the other spin projection. Therefore in the case of a sufficiently long wire the conductance  exhibits a dip to $2e^2/h$ 
where the current is 100\% spin polarized.

To calculate the current through the wire, we use the Landauer - B\"uttiker formula\cite{Buttiker85} for the
zero-temperature total electric conductance
\be
 G = \frac{e^2}{h} \sum\limits_{n_L,n_R, \sigma_L,\sigma_R} |t_{n_R\sigma_R, n_L\sigma_L}|^2
 \label{G}
\ee
where $t_{n_R\sigma_R, n_L\sigma_L}$ are the transmission amplitudes from the state in transverse mode $n_L$ with spin projection 
$\sigma_L$ in the left lead to the state in the mode $n_R$ with spin projection $\sigma_R$ in the right lead. 
The spin conductance $G_z^s = I_s/V$ with respect to the $z$ axis is given by\cite{Zhai05}
\begin{multline}
 G_{z}^s = -\frac{e}{4\pi} \sum\limits_{n_L,n_R, \sigma_L}
 \bigl(
  t^{*}_{n_R\uparrow,n_L\sigma_L} t_{n_R\uparrow,n_L\sigma_L} 
 \\
  - t^{*}_{n_R\downarrow,n_L\sigma_L} t_{n_R\downarrow,n_L\sigma_L}
 \bigr).
 \label{G_s}
\end{multline}

In general, the transmission amplitudes $t_{n_R\sigma_R, n_L\sigma_L}$ can be calculated only numerically. Analytical results may be obtained
for the particular case of strong and nearly constant spin-orbit interaction if one neglects the reflection from the boundary regions where 
the interaction and the confining potential vanish. This is possible if both quantities go to zero in the leads sufficiently smoothly. To make this evident, we perform a unitary transformation of the Hamiltonian with matrix\cite{Levitov03}
\begin{align}
 \hat{S}(x) = \exp[-i\,\hat{\sigma}_z\,\xi(x)/2],
 \nonumber\\
  \xi(x) = \frac{2m}{\hbar^2} \int_{-\infty}^x dx'\,\alpha(x'),
 \label{S}
\end{align}
which eliminates the term linear in $\hat{p}_x$ in it and brings Eqs. (\ref{Schr-multi}) to the form
\begin{subequations}
\label{sys2}
 \begin{align}
   \frac{d^2\bar\psi_1}{dx^2} &+ (m^2\alpha^2\,\hbar^{-4} + \Delta k_1^2)\,\bar{\psi}_1 
   \nonumber\\=&
   -2m\alpha(x)\,p_{12}\,\hbar^{-3} 
   \left( \hat{\sigma}_x\,\cos\xi - \hat{\sigma}_y \sin\xi \right)\,
   \bar\psi_2,
   \\
   \frac{d^2\bar\psi_2}{dx^2} &+ (m^2\alpha^2\,\hbar^{-4} + \Delta k_2^2)\,\bar{\psi}_2 
   \nonumber\\=&
   -2m\alpha(x)\,p_{12}^{*}\,\hbar^{-3} 
   \left( \hat{\sigma}_x\,\cos\xi - \hat{\sigma}_y \sin\xi \right)\,
   \bar\psi_1,
  \end{align}
\end{subequations}
where $\Delta k^2_{1,2} = 2m\,(\eps - \eps_{1,2})/\hbar^2$. Even though $\alpha$ and $U$ are smooth functions of $x$, the right-hand sides 
of equations (\ref{sys2}) contains rapidly oscillating functions $\cos\xi$ and $\sin\xi$ that lead to interband scattering. These
equations are similar to those of mechanical parametric resonance\cite{LL} and can be solved in a similar way. If the detuning in both bands is 
small  and the interband coupling is weak, i.~e. $\Delta k^2_{1,2} \ll m^2\alpha^2/\hbar^4$ and $|p_{12}| \ll m\alpha/\hbar$, the coupled components of the wave function may be presented in the form
\begin{subequations}
\begin{align}
 u_1 = A_1(x)\,e^{i\xi/2} + B_1\,e^{-i\xi/2},
 \\
 v_2 = C_2(x)\,e^{i\xi/2} + D_2\,e^{-i\xi/2},
 \label{ansatz}
\end{align}
\end{subequations}
where $A_1$, $B_1$, $C_2$, and $D_2$
are amplitudes that slowly vary on the scale of $\hbar^2/(m\alpha)$. Substituting Eqs. (\ref{ansatz}) into (\ref{sys2}), neglecting the second derivatives of slowly varying quantities and collecting the terms proportional to $\exp(\pm i\xi/2)$ leads to a system of first-order equations
\begin{subequations}
\label{a1-d2}
\begin{align}
 \frac{2im\alpha}{\hbar^{2}}\,\frac{dA_1}{dx}& = -\Delta k_1^2\,A_1 - \frac{2m\alpha p_{12}}{\hbar^{3}}\,D_2,
 \label{a1}\\
 \frac{2im\alpha}{\hbar^{2}}\,\frac{dB_1}{dx}& =  \Delta k_1^2\,B_1,
 \label{b1}\\
 \frac{2im\alpha}{\hbar^{2}}\,\frac{dC_2}{dx}& = -\Delta k_2^2\,C_2,
 \label{c2}\\
 \frac{2im\alpha}{\hbar^{2}}\,\frac{dD_2}{dx}& =  \Delta k_2^2\,D_2 + \frac{2m\alpha p_{12}^{*}}{\hbar^{3}}\,A_1.
 \label{d2}
\end{align}
\end{subequations}
While the standalone Eqs. (\ref{b1}) and (\ref{c2}) have purely oscillating solutions for any choice of parameters, the solutions of
coupled equations (\ref{a1}) and (\ref{d2}) may exponentially grow or decay. If we assume them to be proportional to $e^{sx}$, one easily finds 
the roots of characteristic equations of system (\ref{a1}) - (\ref{d2})
\begin{align}
 &s_{1,2} = i\hbar^2\,\frac{\Delta k_1^2 - \Delta k_2^2}{4m\alpha} \pm \kappa,
\nonumber\\
 %\kappa = \frac{1}{4m\alpha}\,&\sqrt{ 16m^2\alpha^2|p_{12}|^2 - (\Delta k_1^2 + \Delta k_2^2)^2 }.
 \kappa = &\sqrt{ 
                 \left|\frac{p_{12}}{\hbar}\right|^2 
                 - 
                 \hbar^4 \left(\frac{\Delta k_1^2 + \Delta k_2^2}{4m\alpha}\right)^2
               }.
 \label{roots}
\end{align}
Solving Eqs. (\ref{a1}) - (\ref{d2}) and similar equations for $u_2$ and $v_1$ results in the transmission amplitudes from the left
to the right
\begin{multline}
 |t_{1\uparrow,1\uparrow}|^2 = |t_{2\uparrow,2\uparrow}|^2
\\ =
 \frac
 { 16m^2\alpha^2 |p_{12}|^2 - \hbar^6\,(\Delta k_1^2 + \Delta k_2^2)^2 }
 { 16m^2\alpha^2 |p_{12}|^2\,\cosh^2(\kappa L) - \hbar^6\,(\Delta k_1^2 + \Delta k_2^2)^2 }
 \label{evanesc}
\end{multline}
with
$
 |t_{1\downarrow,1\downarrow}|^2 = |t_{2\downarrow,2\downarrow}|^2 = 1
$
and zero spin-mixing or band-mixing transmission amplitudes. A substitution of these amplitudes into Eqs. (\ref{G}) and (\ref{G_s}) 
suggests that the electric conductance as a function of the Fermi energy has a dip at the second quantization plateau, which
corresponds to a spike in the spin current. This is due to blocking of the current from the left to the right for spin-up electrons
inside the gap in the spectrum. In the strong-interaction approximation, the dip is centered at $\eps = (\eps_1+ \eps_2)/2$ and has a 
width $\Omega =2\alpha|p_{12}|/\hbar$.

\begin{figure}[t]
 \includegraphics[width=8.5cm]{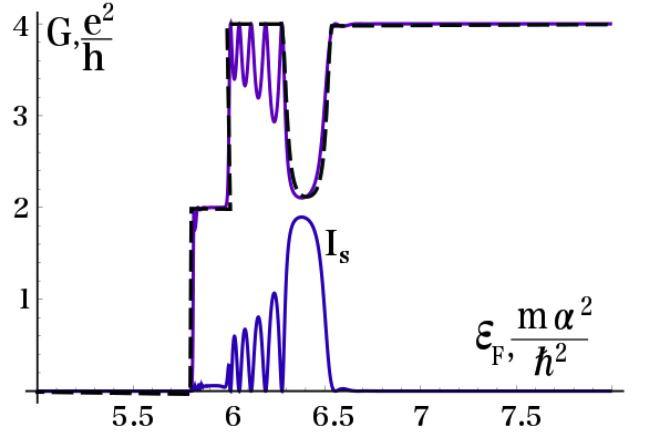}
 \caption{\label{fig4}The dependence of normalized conductance and spin current on the Fermi energy. The upper and lower solid curves
 present $G(\eps)$ and $I_s$ numerically calculated for $|p_{12}| = 0.08m\alpha/\hbar$, $\Delta\eps = 0.32m\alpha^2/\hbar^2$, and $L=20\hbar^2/m\alpha$. 
 The dashed curve shows $G(\eps)$ calculated by means of Eq. (\ref{evanesc}).}
\end{figure}

\begin{figure}[t]
 \includegraphics[width=6cm]{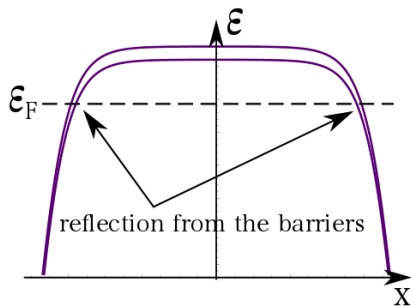}
 \caption{\label{fig5} Spatial dependence of the gap in the spectrum. The lower curve shows the position of the maximum in the lower dispersion curve, and the upper curve shows the position of the minimum in the upper one. Electrons experience partial reflections at the points where the
 Fermi level crosses the gap in the spectrum. }
\end{figure}

Figure \ref{fig4} shows the calculated electric conductance and spin current for $|p_{12}| = 0.08m\alpha/\hbar$, $\Delta\eps = 0.32m\alpha^2/\hbar^2$,
and $L=20\hbar^2/m\alpha$.\bblue{\cite{estimate}} Solid lines show the values obtained by a numerical solution of Eqs. (\ref{sys2}) with account taken of 
spatial variations of $\eps_{1,2}$ and $\alpha$, and the dashed line shows analytical results calculated by means of Eq. (\ref{evanesc}).
Both the numerically calculated conductance and spin current exhibit an oscillatory behavior as the Fermi level approaches the spectrum 
gap from below. This behavior is explained by quantum interference effects that arise due to the reflections of electrons from the ceiling
of the allowed band at the edges of the wire where it goes down (see Fig. \ref{fig5}). The amplitude of the oscillations increases as the gap 
is approached because the reflection amplitude increases.

\section{Conclusion}

We have shown that Rashba spin-orbit interaction may open additional gaps in the spectrum of a multichannel quantum wire if the transverse 
confining potential is chosen appropriately. This happens if the energy levels of transverse quantization come in pairs and the matrix elements of transverse momentum between the corresponding states is sufficiently small. In this case, the conductance of the wire exhibits a dip and the spin current exhibits a spike inside the gap. If the contact is sufficiently long, the conductance in the dip drops from $4e^2/h$ to $2e^2/h$ and the 
current is fully spin-polarized in the transverse in-plane direction. This effect may be used for designing an all-electrical spin transistor. By
applying a negative voltage to the middle longitudinal gate, one may increase the degree of spin polarization of the current from zero to 100\% if 
the Fermi level is adjusted appropriately.

\bblue{One of the main advantages of using electric bias for spin control
is the ability to make it time-dependent. This can lead to non-trivial
effects in the transport.\cite{Sadreev13} In the future, it would be of interest to study the effects of time-periodic bias in our model.}

\begin{acknowledgments}
This work was supported by Russian Foundation for Basic Research, grant 13-02-01238-a, and by the program of Russian Academy of Sciences.
\end{acknowledgments}

\end{document}